\documentclass[preprint,12pt]{elsarticle}

\usepackage{amssymb}
\usepackage{amsmath,bm}
\usepackage{mathtools}
\usepackage{amsbsy}
\usepackage{amssymb}
\usepackage{amscd}
\usepackage{amsfonts}
\usepackage{array}
\usepackage{graphics}
\usepackage{subfigure}
\usepackage{xcolor}
\usepackage{nicefrac}
\usepackage{color}
\usepackage{lineno}
\usepackage[margin=0.8in]{geometry}
\usepackage{gensymb}

\newcommand{\mbs}[1]{\boldsymbol{#1}}

  \def\bF{{\mbs{F}}}
  
 \def\bK{{\mbs{K}}} 
  
  \def\bR{{\mbs{R}}}

 \def\b0{{\mbs{0}}}

\def\ba{{\mbs{a}}}  
 \def\be{{\mbs{e}}}

  \def\bu{{\mbs{u}}}

\def \Tet {\mbs{\Theta}}

\def    \obu {{\overline{\bu}}}

\def    \obK {{\overline{\bK}}}

\def    \hbu {{\hat{\bu}_i}}
\def    \hbv {{\hat{\bu}_j}}

\journal{International Journal of Engineering Science}

\begin{document}

\begin{frontmatter}

\title{A numerical model of the human cornea accounting for the fiber-distributed collagen microstructure}

    \author[label1]{Maria Laura De Bellis}
    \author[label1]{Marcello Vasta}
    \author[label2]{Alessio Gizzi}
    \author[label3]{Anna Pandolfi\corref{cor1}}

    \address[label1]{
    Dipartimento INGEO,
    Universit\`a di Chieti-Pescara,
    Viale Pindaro 42,
    Pescara, Italy}
    
    \address[label2]{
    Department of Engineering,
    Universit\'a Campus Bio-Medico di Roma,
    Via A. del Portillo 21, 00128 Rome, Italy\\
    Research Unit of Nonlinear Physics and Mathematical Modeling}
    %\cortext[cor1]{Corresponding author: a.gizzi@unicampus.it}

    \address[label3]{
    Politecnico di Milano,
    Dipartimento di Ingegneria Civile ed Ambientale,
    Piazza Leonardo da Vinci 32,
    Milano, Italy
    \cortext[cor1]{Corresponding author: {anna.pandolfi@polimi.it}}}


%\author[inst1]{Author One}

%\affiliation[inst1]{organization={Department One},%Department and Organization
%            addressline={Address One}, 
%            city={City One},
%            postcode={00000}, 
%            state={State One},
%            country={Country One}}

%\author[inst2]{Author Two}
%\author[inst1,inst2]{Author Three}

%\affiliation[inst2]{organization={Department Two},%Department and Organization
%            addressline={Address Two}, 
%            city={City Two},
%            postcode={22222}, 
%            state={State Two},
%            country={Country Two}}

\date{\today}

\begin{abstract}
We present a fiber-distributed model of the reinforcing collagen of the human cornea. The model describes the basic connections between the components of the tissue by defining an elementary block (cell) and upscaling it to the physical size of the cornea. The cell is defined by two sets of collagen fibrils running in sub-orthogonal directions, characterized by a random distribution of the spatial orientation and connected by chemical bonds of two kinds. The bonds of the first kind describe the lamellar crosslinks, forming the ribbon-like lamellae; while the bonds of the second kind describe the stacking crosslinks, piling up the lamellae to form the structure of the stroma. The spatial replication of the cell produces a truss structure with a considerable number of degrees of freedom. The statistical characterization of the collagen fibrils leads to a mechanical model that reacts to the action of the deterministic intraocular pressure with a stochastic distribution of the displacements, here characterized by their mean value and variance. The strategy to address the solution of the heavy resulting numerical problem is to use the so-called \emph{stochastic finite element improved perturbation method} combined with a fully explicit solver. Results demonstrate that the variability of the mechanical properties affects in a non-negligible manner the expected response of the structure to the physiological action.

\end{abstract}


%%\item Research highlight 2

\begin{keyword}
Fibers distributed \sep 
Cornea microstructure \sep 
Collagen stiffness \sep
Stochastic finite element \sep
Improved perturbation method
\end{keyword}

\end{frontmatter}

\date{\today}

%\linenumbers

\section{Introduction}
\label{sec:introduction}

The human cornea, a transparent and relatively stiff tissue, is the convex-concave external lens of the eye. The spherical shape provides two-thirds of the total refractive power of the eye and offers confinement to the internal fluids. The intraocular pressure (IOP) acts on the posterior surface of the cornea and induces an overall tensile stress state in the tissue. The cornea is organized into several layers: the central stroma, the thickest layer, carries the structural functions. As in many biological tissues with structural functions, the main component of the stroma is the collagen, organized hierarchically in microstructures at different scales, including the fibril filaments up to the ribbon-like lamella. X-ray imaging has revealed that the preservation of the spherical shape and the correct light refraction of the cornea is due to the particular architecture of the lamellae. Most collagen fibrils are randomly distributed, as necessary to provide uniform protection, but about 1/3 of the stromal collagen is clearly oriented in the nasal-temporal (NT) and in the superior-inferior (SI) directions to sustain extra loads due to the activity of the eyelids and the eye muscles~\cite{meek:2009}. Another region where collagen fibrils run in a specific (circumferential) direction is the limbus, the periphery annulus that links the cornea to the sclera and to the iris~\cite{kokott1938mechanisch}.   

Alterations of the stromal microstructure may induce modifications of the regular shape of the cornea. Among others, the pathology known as keratoconus refers to a localized loss of the collagen organization, causing thinning and reshaping of the cornea into a non-physiological \emph{conus}: the associated effects on vision are irregular astigmatism, myopia, and strong aberrations. Despite numerous experimental and theoretical investigations, the etiology of keratoconus has yet to be clarified. However, recent discrete models of the collagen microstructure of the cornea, including lamellar collagen and chemical bonds disposed to form a representative cell \cite{pandolfi:2019}, have been able to describe the healthy and the keratoconus geometries of the human cornea by assuming suitable distributions of the mechanical properties of the structure components, corresponding to the healthy and the diseased tissue, respectively. Subsequent works have adopted an ansatz about the spatial and temporal evolution of the reduction of the mechanical property by considering the diffusion of a damage-like parameter~\cite{Gizzi:2018aa}. Very realistic keratoconus shapes were obtained by introducing a constitutive law for the chemical bonds based on a damaging chemical potential \cite{pandolfi2022modeling}.

These studies were conducted under deterministic assumptions. Still, it is well known that, in biological tissues, uncertainties related to the micro-architecture and to the mechanical properties play a critical role and cannot be neglected. The stochastic aspects of the mechanics of the cornea require the adoption of alternative numerical approaches, able to include the relevant uncertainties and provide a numerical response that accounts for the variability in terms of displacement and stress distributions.

A well-established approach used to characterize the response of structures with uncertain parameters from a probabilistic point of view is the Stochastic Finite Element Method \cite{stefanou2009stochastic, pryse2018projection,falsone2013explicit, navarro2022new}. Among many other solutions procedures available in literature \cite{ghanem2003stochastic}, perturbation approaches have been proven to be more sound, and feasible \cite{liu1986probabilistic, kleiber1992stochastic}. The key features of perturbation approaches are (i) the coupling of classical finite element methods and perturbation techniques and (ii) the Taylor expansion of random variables characterized by a prescribed zero-mean stochastic distribution law. Perturbation approaches deploy the structural response in terms of mean and covariance functions, obtained as first-order and second-order moments of the attendant variables.

Notoriously, perturbation approaches applied to solve SFEM lose robustness as the uncertainty of the variables increases. A more accurate statistical description of the stochastic variables is obtained through the ``improved perturbation method'', initially proposed for linear static and dynamic structural problems \cite{elishakoff1995improved, muscolino2000improved}, later extended to geometrically nonlinear problems \cite{impollonia2002static}, and also applied to more general cases \cite{van2003modal, kaminski2013stochastic, kasinos2021reduced}. The improved perturbation method is a first-order perturbation technique, where the ``improved'' mean and covariance of the stochastic variables are evaluated by accounting for information on the mean and correlation between uncertain parameters. 

In the present study, we apply the improved perturbation method to the discrete model of the cornea proposed in \cite{pandolfi:2019}, to introduce in a proper mathematical way the intrinsic stochastic features of the micro-components of the human stroma. To obtain a numerically tractable problem, we reformulate the approach in terms of internal and external force balance and solve it explicitly. The approach is able to provide the mean array and the covariance matrix of the displacements of the discrete model of the cornea undergoing the IOP action.

The paper is organized as follows. In Section~\ref{sec:geometricModel}, we describe the micro-mechanical model of the cornea. Specifically, in Section~\ref{sec:stiffMatrix}, we define the stochastic stiffness matrix, which accounts for the spatially distributed collagen and is responsible for the local variability of the structural element stiffness. In Section~\ref{sec:statApprox}, we describe the statistical approximation of the equilibrium of the discrete model via the improved perturbation method; Section~\ref{sec:numerics} discusses the numerical strategy used to solve the resulting equation systems. In Section~\ref{sec4}, we provide a quantitative characterization of the proposed statistical model in terms of mean, standard deviation, and coefficient of variation for the optical axis displacements. Finally, in Section \ref{sec:fR}, we recollect the main achievements of the study and discuss possible future extensions and applications.

\section{A fiber distributed model of the corneal microstructure}
\label{sec:geometricModel}

\begin{figure}[t]
	\centering
    \includegraphics[width=\linewidth]{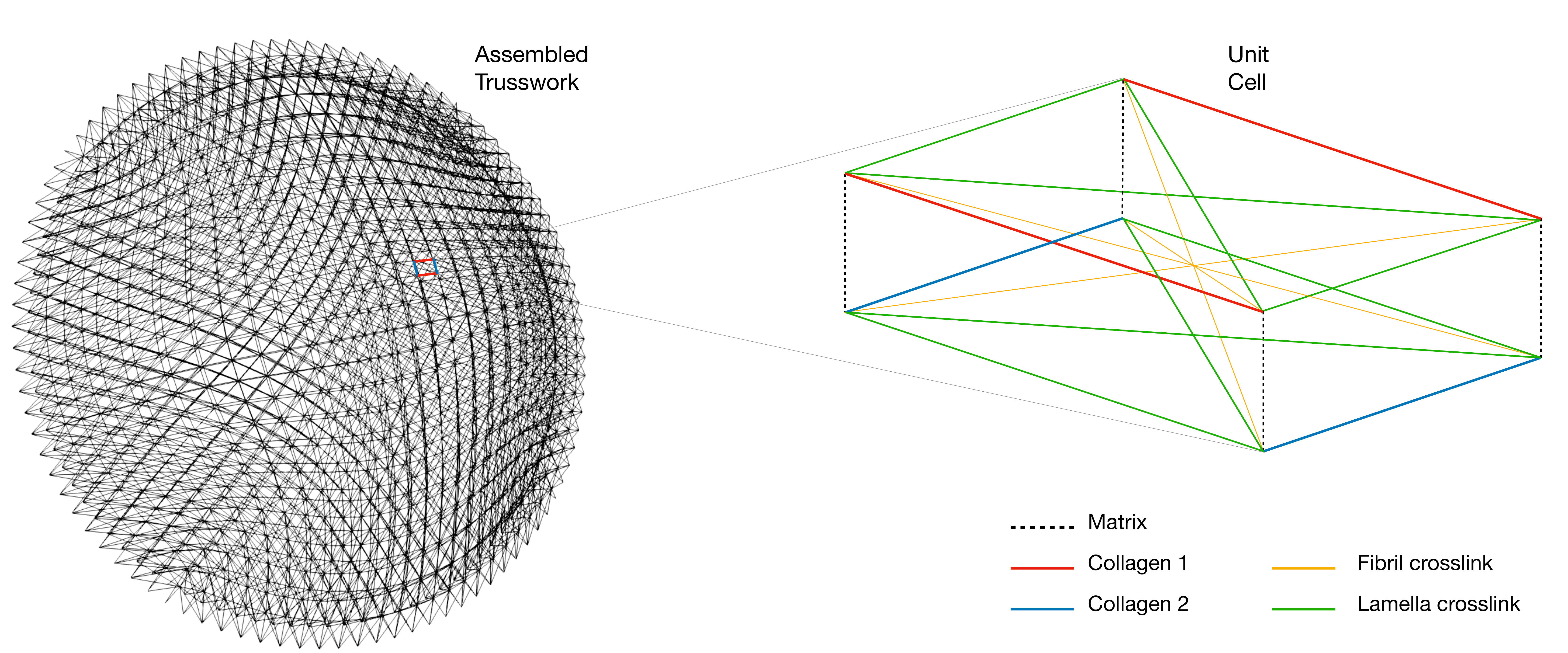}
	\caption{Assembled trusswork model (left) and representative unit cell (right) highlighting the different truss elements: 
        extracellular matrix (dashed black),
        collagen fibers (red-blue),
        fibril crosslink (orange),
        lamella crosslink (green).}
	\label{fig1:geom}
\end{figure}

The concept of a discrete model of the structure of the human cornea proposed in \cite{pandolfi:2019} will be taken as the starting point for the present study. A deterministic truss network models, at the macro-scale, the idealized fundamental structural unit of the stroma, consisting of two sets of parallel collagen fibrils superposed orthogonally and kept in their configuration by the presence of chemical bonds (crosslinks) and with elastic separators (extracellular matrix). The chemical bonds act to form a lamellar sheet and to stack, one on top of the other, two lamellae of different orientations, see Fig.~\ref{fig1:geom} (right).

For the sake of clarity, we briefly recall the features of the trusswork as it was originally conceive, i.~e., a deterministic structure. Then we show how we enhance the model by accounting for the clinically observed dispersion of the collagen fibrils forming the lamellae.

\subsection{The underlying geometry}
\label{ssec:underlyingGeometry}

The orientation of the trusses in the model has been chosen to reflect the dominant direction of the collagen fibrils averaged across the thickness. The main orientation of the fibrils gradually varies from the orthogonal arrangement at the center, which follows the NT and SI directions, to the orthogonal arrangement at the limbus, where fibrils run circumferentially and radially. The approximation disregards the variation of the collagen distribution from the anterior to the posterior surfaces of the cornea, which is a minor issue that can be easily fixed.

Following the geometry of an idealized cornea with average sizes (the extension to a patient-specific model is trivial), we use the geometric parameters listed in Table~\ref{table:geometry}, which are easily obtained by a clinical measurement. The parameters refer to a cornea deformed by the IOP. Note that, for numerical reasons (stability of the structure, which comprises only the anterior and the posterior layer), we need to reduce the central corneal thickness to an unrealistic value of 0.3~mm, against an average of 0.6~mm for healthy human corneas. One discretization of the geometry used in the study is shown in Fig.~\ref{fig1:geom} (left).

\begin{table}[]
    \centering
    \begin{tabular}{llrc}
                           & Parameter                &  Value    &  Unit \\
         \hline \hline
         Apex elevation    &                           & 2.54  & mm \\
         Thickness         & Center                    & 0.30  & mm \\
         Anterior surface  & Steepest meridian radius  & 7.52  &  mm\\
                           & Flattest meridian radius  & 7.90  &  mm\\
                           & Asphericity coefficient   & -0.15 &  mm\\
         Posterior surface & Steepest meridian radius  & 6.77  &  mm\\
                           & Flattest meridian radius  & 6.90  &  mm\\
                           & Asphericity coefficient   & -0.15 &  mm\\
         In-plane diameter &                           & 11.24 &  mm \\
         In-plane orientation & Steepest meridian      & 128   &  deg \\
                              & Flattest meridian      & 38   &  deg \\
         \hline
    \end{tabular}
    \caption{Geometric parameters of the physiologic cornea}
    \label{table:geometry}
\end{table}

The assembled network results from the spatial repetition of a unit cell, as illustrated in Fig.~\ref{fig1:geom}(right). The truss elements modeling the collagen fibrils (blue and red) are arranged on two parallel hyperplanes. Blue trusses model a lamella on the posterior surface, red trusses model a lamella on the anterior surface, and they run orthogonally, one with respect to the other. Green trusses model lamellar crosslinks that confer the ribbon-like shape to the lamella; yellow trusses model the crosslinks that bond the two lamellae to form a double shell. Fig.~\ref{fig1:geom}(left) shows that the orientation of the unit cell switches smoothly from NT-SI at the center to circumferential-radial at the periphery. The assembled network results in a mechanically stable structure if the inclination of the yellow trusses is around 45 degrees, which requires a reduction of the thickness for finer discretizations. This drawback, related to the model simplicity, has been addressed in a parallel study \cite{koery:2023}. Furthermore, we have the possibility to introduce additional trusses (broken line black) orthogonal to both lamellae, which model the action of the extracellular matrix surrounding the collagen fibrils in the stroma. However, this feature has not been used in the current study.

The IOP action is described in terms of point loads applied to the posterior surface of the cornea in the direction normal to the surface, as done in \cite{pandolfi:2019}. The truss nodes resting on the model external annulus, corresponding to the limbus, are considered fixed. For future reference, we remark that the unit cell is defective at the limbus, because some of the composing elements (either collagen or lamellar crosslink) are missing and the diagonal crosslinks are longer. The effect of this heterogeneity in the discretization, visible in all the images reporting numerical results, is to confer to the boundary annulus of the shell a larger compliance in the radial direction and a larger (infinite) stiffness in the circumferential direction. For this reason, deterministic and stochastic displacements at the limbus are affected by artifacts. Clearly, this effect vanishes with the reduction of the discretization size. 

All trusses (fibrils and crosslinks) are modeled as classical linear elastic trusses. The $i$-th truss is characterized by the stiffness $k_i = E_i A_i/h_i$, where  $A_i$ is the equivalent cross-section area, $h_i$ is the length of the truss and $E_i$ is the equivalent Young modulus, obtained by exploiting a properly conceived homogenization procedure based on the experimental results available in the literature \cite{marino:2013, gyi:1988, scott:2003}, as discussed in \cite{pandolfi:2019}.

\subsection{Statistical characterization of the collagen trusses}
\label{sec:stiffMatrix}

Next, we enrich the model by assuming that the orientation of the collagen trusses is characterized by a statistical distribution that describes the spatial dispersion of the fibrils. The trusses modeling the crosslinks, by contrast, are kept deterministic.

The inclusion of the collagen fibril distribution statistics is achieved by the improved perturbation stochastic approach, developed for the solution of linear elastic problems \cite{elishakoff1995improved} and later extended to nonlinear problems \cite{impollonia2002static}. In the present study, we found that the nonlinear formulation is more appropriate.
 
\begin{figure}[h]
	\centering
	\includegraphics[width=\linewidth]{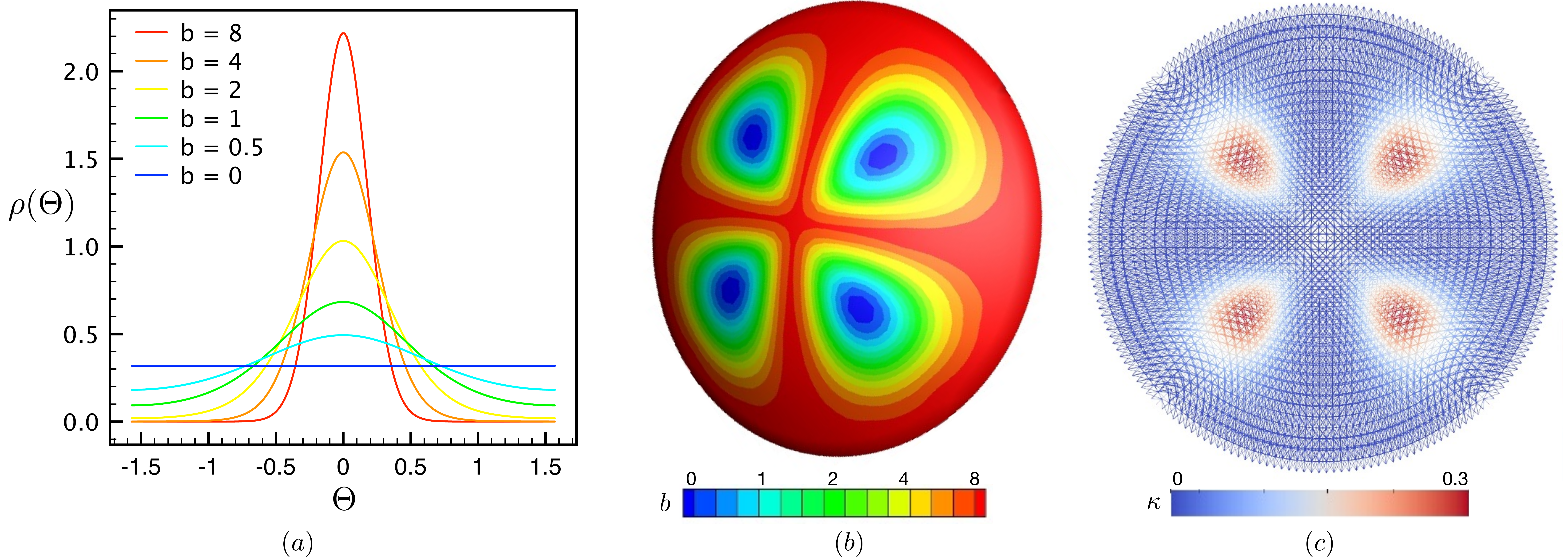}
	\caption{(a) Aspect of the von Mises PDF, for a few significant values of the concentration parameter $b\in[0\div8]$. (b) Three-dimensional continuum model of the human cornea, showing the contour levels of the assumed spatial distribution of the concentration parameter $b$, cf.~\cite{pandolfi2012fiber}. (c) Trusswork model adopted in the stochastic analysis. Top view of maps of the integral coefficient $\kappa$, Eq.~\eqref{eq:9}.}
	\label{fig2:vonMises}
\end{figure}

We begin by assuming that the deterministic orientation of each collagen truss (blue and red elements in Fig.~\ref{fig1:geom}) corresponds to the average orientation of a bunch of fibrils randomly oriented. We denote the average orientation of a collagen truss by $\ba_0$ and assume that the generic collagen fibril orientation $\ba$ follows a specified symmetric probability distribution function (PDF) $\rho(\ba)=\rho(-\ba)$, with rotational symmetry about $\ba_0$, cf.~\cite{pandolfi2012fiber, vasta2014three}. The rotational symmetry of the PDF, which leads to a transversely isotropic response for the material, requires that the PDF in spherical coordinates depends only on the polar (meridian) angle $\Theta \in \left[0,\pi \right]$ and not on the azimuthal angle $\Phi \in \left[0,2\pi \right]$. At a generic point, the integral over a unit sphere $\omega$ reduces to
\begin{linenomath}
\begin{align}\label{eq:1}
   \frac{1}{4 \pi} \int_{\omega} \rho(\ba) d \omega = 1 \, ,
\end{align}
\end{linenomath}
and assuming $\rho(\ba)=\rho(\Theta)$, recalling that $d \omega= \sin{\Theta} d \Theta d \Phi$, it follows that
\begin{linenomath}
\begin{align}
    \int_0^{\pi}  \rho(\Theta) \sin{\Theta}  d \Theta = 2.
\end{align}
\end{linenomath}
For the chosen fiber distribution and generic function $f$, the operator $\langle \cdot \rangle$ returns the average of $f$ over the unit sphere as
\begin{linenomath}
\begin{align}
 \langle f \rangle
	= \dfrac{1}{4\pi}
	\int_0^{2\pi} d\Phi
	\int_0^\pi \rho( \Theta) f \sin \Theta d \Theta=\dfrac{1}{2} \int_0^\pi \rho( \Theta) f \sin \Theta d \Theta.
    \label{eq:average}
\end{align}
\end{linenomath}
Following \cite{pandolfi2012fiber, vasta2014three}, we assume here that the density function $\rho (\Theta)$ is characterized by a $\pi$-periodic normalized von Mises distribution of the form
\begin{linenomath}
\begin{equation}
	\rho( \Theta) = \dfrac{1}{ I} e^{b\cos(2 \Theta)} \,,\quad
	I = \dfrac{1}{\pi} \int_0^\pi e^{b\cos(2 \Theta)} d \Theta \,,\label{eq:5}
\end{equation}
\end{linenomath}
where $b$ is the concentration parameter, which controls how the distribution arranges around the mean value, see Fig.~\ref{fig2:vonMises}(a). It follows that the average of $\Theta^2$ is given by the expression
$$ 
	\langle \Theta^2 \rangle = \dfrac{1}{2} \int_0^\pi \rho( \Theta) \Theta^2 \sin \Theta d \Theta =
	\dfrac{1}{2 I} \int_0^\pi  e^{b\cos(2 \Theta)} \Theta^2 \sin \Theta d \Theta	\, .
$$
For an assigned distribution, we also introduce the dispersion parameter $\kappa$ defined as \cite{pandolfi2012fiber}:
\begin{linenomath}
\begin{equation}
	\kappa
	=
	\dfrac{1}{4} \int_0^\pi \rho( \Theta) \sin^3 \Theta d \Theta \, , 
   \label{eq:9}
\end{equation}
\end{linenomath}
which plays the inverse role of $b$, i.e., $k$ increases for low $b$, and represents a measure of the degree of dispersion of the fibers.

In previous studies, considering continuum models of the cornea, we accounted for the variability of the dispersion of the collagen fibrils by adopting a map of the $b$ parameter characterized by a double symmetry with respect to the SI and NT directions, cf.~\cite{pandolfi2012fiber}. Figure \ref{fig2:vonMises}(b) shows the maps of $b$ over the solid model of the cornea. Interestingly, high values of $b$ (describing a strong alignment of the collagen fibrils about the mean orientation and thus an anisotropic behavior of the tissue) are observed at the center, forming a SI and NT direction cross, and all along the circular limbus. The low $b$ zones are disposed to form four lobes, where the collagen fibrils are more dispersed, conferring a quasi-isotropic behavior to the tissue. We adopt the same maps of $b$ also for the trusswork model.

By referring to a network model of the collagen reinforcement, we now consider every collagen truss $e$ as representative of the local distribution of fibrils, characterized by the concentration coefficient $b$ obtained from the map in Fig.~\ref{fig2:vonMises}(b).

To compute the equivalent truss stiffness of the distribution, we recall that the 3D stiffness matrix $\bK_e$ of the truss element $e$ referred to its intrinsic frame (i.~e., $\ba_0$ aligned with the basis axis $\be_1$) is 
\begin{linenomath}
\begin{equation}
   \widehat{\bK}_e
   =
   \dfrac{E_e A_e}{h_e}
	   \begin{bmatrix}
		  1 & 0 & 0 & -1 & 0 & 0\\
		  0 & 0 & 0 &  0 & 0 & 0\\
		  0 & 0 & 0 &  0 & 0 & 0\\
		 -1 & 0 & 0 &  1 & 0 & 0\\
		  0 & 0 & 0 &  0 & 0 & 0\\
		  0 & 0 & 0 &  0 & 0 & 0 
	\end{bmatrix}
    \, ,
    \label{truss1}
\end{equation}
\end{linenomath}
where $E_e$ is the elastic modulus of the material, $A_e$ is the cross-section area, $h_e$ the axis length. The three quantities should refer to the individual collagen filament, but, in the present context, following \cite{pandolfi:2019}, we will resort to equivalent homogenized values.

For a generic collagen fibril belonging to the axis-symmetric distribution and inclined of an angle $\Theta$ with respect to the main orientation $\ba_0=\be_1$, the components of the stiffness matrix $\bK_e$ become
\begin{linenomath}
\begin{equation}
	\bK_e  
    = 
    {\bR_e}^T \widehat{\bK}_e \bR_e
    =
    \dfrac{E_e A_e}{h_e}
	   \begin{bmatrix}
		 c^2 & cs   & 0 & -c^2 & -cs  & 0\\
		  cs & s^2  & 0 &  -cs & -s^2 & 0\\
		   0 &   0  & 0 &    0 &    0 & 0\\
		-c^2 & -cs  & 0 &  c^2 &  cs  & 0\\
		 -cs & -s^2 & 0 &   cs &  s^2 & 0\\
		   0 &    0 & 0 &    0 &    0 & 0
	\end{bmatrix}\label{truss2} \,,
\end{equation}
\end{linenomath}
where $s=\sin \Theta, c=\cos \Theta$ and $\bR_e$ is the rotation matrix
\begin{linenomath}
\begin{equation}\label{eq:rotation}
    \bR_e   
    =
    \begin{bmatrix}
     c & s &  0 &  0 & 0 & 0 \\
    -s & c &  0 &  0 & 0 & 0 \\
     0 & 0 &  1 &  0 & 0 & 0 \\
     0 & 0 &  0 &  c & s & 0 \\
	   0 & 0 &  0 & -s & c & 0 \\
     0 & 0 &  0 &  0 & 0 & 1 
    \end{bmatrix} \,.
\end{equation}
\end{linenomath}
The average stiffness matrix expressed in the intrinsic reference frame associated with $\ba_0$ reads
\begin{linenomath}
\begin{equation}\label{eq:AVGstiff}
	\langle \widehat{\bK}_e \rangle = \dfrac{E_e A_e}{h_e}
	\begin{bmatrix}
		1-2\kappa & 0 & 0 & 2\kappa-1 & 0 & 0 \\
		0 & 2\kappa & 0 & 0 & -2\kappa & 0\\
        0 & 0 & 0 & 0 & 0 & 0\\
		2\kappa-1 & 0 & 0 & 1-2\kappa & 0 & 0 \\
		0 & -2\kappa & 0 & 0 & 2\kappa & 0\\
        0 & 0 & 0 & 0 & 0 & 0
	\end{bmatrix} \,,
\end{equation}
\end{linenomath}
where we make use of Eq.~\eqref{eq:9} for the expected value of the non-zero terms in Eq.~\eqref{truss2}
\begin{linenomath}
$$\
    \langle c^2 \rangle
	=
	1 - 2 \kappa
	\, \qquad
	\langle cs \rangle
	=
	0
	\, , \qquad
	\langle s^2 \rangle
	=
	2\kappa	\, .
$$
\end{linenomath}
The dispersion parameter $\kappa$ varies within the shell geometry,  reflecting the distribution of $b$. The map of $\kappa$ over the discrete network is visualized in Fig.~\ref{fig2:vonMises}(c).

The average stiffness matrix in Eq.~\eqref{eq:AVGstiff} corresponds to an ``enhanced truss element'' since it is characterized by the presence of additional terms. Consistently, in the limit of $\kappa \rightarrow 0$, i.~e., when the degree of dispersion becomes negligible, the enhanced model converges to the deterministic model of the truss, see Eq.~\eqref{truss1}. We note that, while the truss length $h_e$ and the truss area $A_e = t \Delta x$ in Eq.~\eqref{eq:AVGstiff} are defined by the actual discretization, and therefore are linked to the mesh size $\Delta x$, as well as to the cornea thickness $t$, for the equivalent elastic modulus of the healthy collagen we will use $E_e=0.15$ MPa, as done in \cite{pandolfi:2019}.

%%%%%%%%%%%%%%%%
\subsection{Statistical approximation of equilibrium}
\label{sec:statApprox}

Next, we state the equilibrium conditions for a stochastic trusswork made of $N$ enhanced trusses modeling the collagen under the action of deterministic forces. We characterize the probabilistic nature of the problem by introducing stochastic angles $\Theta_i$, $i=1,...,N$, that define the magnitude of the inclination of the truss axes with respect to their deterministic orientation ${\ba_i}_0$. We assume the $\Theta_i$ to be statistically independent (i.~e., uncorrelated) and symmetric around its zero mean and write
\begin{linenomath}
\begin{equation}\label{eq:stat}
	\langle \Theta_i \rangle = 0 \,,\quad
	\langle \Theta_i \Theta_j \rangle = \langle \Theta_i^2 \rangle \delta_{ij}, \,,\quad \langle \Theta_i \Theta_j \Theta_k \rangle =0,
\end{equation}
\end{linenomath}
where $\delta_{ij}$ is the Kronecker delta.

Let us consider the discretized structure as generally nonlinear. The size of the unknown stochastic displacement array $\bu$ equals the number $3n$ of degrees of freedom of the problem, where $n$ is the number of nodes. The balance equation in the static regime reads
\begin{linenomath}
\begin{equation}\label{33}
    \bF_{\rm int} \left( \bu ( \Tet ), \Tet \right) 
    = 
    \bF_{\rm ext} \, ,
\end{equation}
\end{linenomath}
where the external forces depend on the IOP and the internal forces depend on the displacements, and both internal forces and displacements depend on the array $\Tet$ collecting the $N$ random variables $\Theta_i$. In order to exploit the proposed procedure, let us now Taylor expand about $\Tet=\b0$ and up to the second order terms the internal forces, seen as functions of the uncertain parameters, as
\begin{linenomath}
\begin{equation}\label{34}
    \bF_{\rm int} \left( \bu ( \Tet ), \Tet \right)
    \approx 
    \left. \bF_{\rm int} \right|_{\Tet=\b0} 
    +
    \left. \frac{\partial \bF_{\rm int}}{\partial \Theta_i} \right|_{\Tet=\b0} 
    \Theta_i
    +
    \frac{1}{2} \left.
    \frac{\partial^2 \bF_{\rm int}}{\partial \Theta_i \partial \Theta_j} \right|_{\Tet=\b0}
    \Theta_i \Theta_j \,.
\end{equation}
\end{linenomath}

For the sake of simplicity, consistently with \cite{impollonia:2002}, we first consider a linear dependence of the displacement vector on the random variables, and introduce the linear expansion of the displacement array around the mean value $\obu$ as
\begin{linenomath}
\begin{equation}\label{35}
    \bu (\Tet) = 
    \obu + \Theta_i \, \hbu \,,
\end{equation}
\end{linenomath}
where we implicitly assume sum over the index $i=1,..., N$ and $\hbu$ (denoting the first derivative of the displacement field with respect to $\Theta_i$) being an array with size 3$n$, where only the six components corresponding to the two end nodes of the $i$-th enhanced truss are nonzero. Thus, $\bu=\obu$ when $\Tet = \b0$.
We write the internal forces as
\begin{linenomath}
\begin{equation}\label{36}
    \bF_{\rm int} \left( \bu (\Tet ), \Tet \right)
    :=
    \bF_{\rm int} \left( \obu + \Theta_i \hbu, \Tet \right)  \,,
\end{equation}
\end{linenomath}
to point out the double dependence of the internal forces array on the stochastic variable array $\Tet$, a dependence that must be accounted for in the subsequent differentiation procedures. 

We write Eq.~\eqref{34} as
\begin{linenomath}
\begin{equation}\label{37}
    \bF_{\rm int} \left( \obu + \Theta_i \hbu, \Tet \right) 
    =
    \bF^0_{\rm int} \left( \obu \right)
    +
    \bF_{{\rm int},i} \left( \obu, \hbu \right) \Theta_i
    +
    \bF_{{\rm int},ij} \left( \obu, \hbu, \hbv \right)
    \Theta_i \Theta_j \,,
\end{equation}
\end{linenomath}
where we define
\begin{linenomath}
\begin{equation}\label{38}
    \bF^0_{\rm int} (\obu) = \bF_{\rm int} \left( \obu , \b0 \right) \, .
\end{equation}
\end{linenomath}

Next we assume a linear dependence of the nodal forces on the displacements
\begin{linenomath}
\begin{equation}\label{39}
    \bF_{\rm int}(\obu + \Theta_i \hbu, \Tet) = \bK (\Tet)  
    \left(\obu +\Theta_i \hbu\right)  \,,
\end{equation}
\end{linenomath}
where $\bK(\Tet)$ denotes the network tangent stiffness ($3n \times 3n$). Thus, the constant term in Eq.~\eqref{37} becomes
\begin{linenomath}
\begin{equation}\label{40}
    \bF^0_{\rm int} (\obu) = \bK(\b0) \obu \, ,
\end{equation}
\end{linenomath}
and the coefficient of the linear term in Eq.~\eqref{37} can be rendered as
\begin{linenomath}
\begin{equation}\label{41}
    \bF_{{\rm int},i} (\obu, \hbu) =
    \left[ 
    \frac{\partial \bF_{\rm int}}{\partial \Theta_i}
    +
    \frac{\partial \bF_{\rm int}}{\partial \bu}
    \frac{\partial \bu}{\partial \Theta_i}
    \right]_{\Tet=\b0} \, .
\end{equation}
\end{linenomath}

By accounting for Eq.~\eqref{39}, we obtain
\begin{linenomath}
\begin{equation}\label{42}
    \bF_{{\rm int},i} (\obu, \hbu) =
    \left[ \frac{\partial \bK (\Tet)}{\partial \Theta_i} \left(\obu +\Theta_i \hbu\right) 
    +
    \bK \left( \Tet \right) \hbu 
    \right]_{\Tet=\b0} = 
    \bK_i \obu + \obK \hbu \, ,
\end{equation}
\end{linenomath}
where we introduced
\begin{linenomath}
\begin{equation}\label{43}
    \bK_i 
    = 
   \left[ \frac{\partial \bK (\Tet)}{\partial \Theta_i} \right]_{\Tet=\b0}\, , \qquad
    \obK = \bK \left( \b0 \right)
    =
    \frac{\partial \bF_{\rm int}}{\partial \bu} \, .
\end{equation}
\end{linenomath}
Here, $\bK_i$ is the derivative of the stiffness matrix with respect to the stochastic variable $\Theta_i$, where only 36 coefficients, disposed in 3$\times$3 blocks, are non-zero. The term $\obK$ is the averaged stiffness matrix, collecting the enhanced truss average stiffness defined in Eq.~(\ref{eq:AVGstiff}). The coefficient of the second order term in Eq.~\eqref{37} is
\begin{linenomath}
\begin{equation}\label{44}
    \bF_{{\rm int},ij} \left( \obu, \hbu\right) =
    \frac{1}{2} 
    \left[
    \frac{\partial}{\partial \Theta_j}
    \left( \frac{\partial \bK (\Tet)}{\partial \Theta_i} \left(\obu +\Theta_i \hbu\right) +
    \bK \left( \Tet \right) \hbu \right) \right]_{\Tet=\b0} \, ,
\end{equation}
\end{linenomath}
in compact form
\begin{linenomath}
\begin{equation}\label{47}
    \bF_{int,ij} \left( \obu, \hbu\right) =
    \frac{1}{2} 
  \left(  \bK_{ij} \obu + \bK_i \delta_{ij}\hbu+ \bK_j \hbu \right)\, ,
\end{equation}
\end{linenomath}
%\begin{linenomath}
%\begin{equation}\label{45}
%    \bF_{{\rm int},ii} \left( \obu, \hbu\right) =
%    \frac{1}{2}
%    \left(
%    \frac{\partial^2 \bK(\Tet)}{\partial \Theta_i^2 } \obu
%    + 
%(    2 \frac{\partial \obK}%{\partial \Theta_i} \hbu  \right) \, .
%\end{equation}
%\end{linenomath}
where  we introduce
\begin{linenomath}
\begin{equation}\label{46}
    \bK_{ij} 
    = 
    \left[\frac{\partial^2 \bK (\Tet)}{\partial \Theta_i \partial\Theta_j} \right]_{\Tet=\b0}.
\end{equation}
\end{linenomath}

By combining Eqs.~\eqref{40}-\eqref{42}-\eqref{47}, the discrete equilibrium equation in \eqref{33} reads
\begin{linenomath}
\begin{equation}\label{48}
    \bF_{\rm ext} =
    \obK \obu 
    +
    \left( \bK_i \obu 
    +
    \obK
    \hbu \right) \Theta_i
    +
     \frac{1}{2} \left( 
    \bK_{ij} \obu + \bK_i  \delta_{ij}\hbu + \bK_j \hbu
    \right) \Theta_i \Theta_j \, .
\end{equation}
\end{linenomath}

The improved perturbation method leads to $N + 1$ sets of equations as in \cite{impollonia:2002}. The first set of equations is obtained by applying the average operator \eqref{eq:average} to both sides of the balance equation \eqref{48}
\begin{linenomath}
\begin{equation}\label{49}
    \bF_{\rm ext} =
    \obK \obu 
    +
    \left(
    \frac{1}{2} 
    \bK_{ii} \obu + \bK_i \hbu 
    \right) \langle \Theta_i^2 \rangle  \, ,
\end{equation}
\end{linenomath}
and the other $N$ sets of equations are given multiplying both sides of Eq.~\eqref{48} by $\Theta_i$ and applying the average operator 
\begin{linenomath}
\begin{equation}\label{50}
    \bK_i \obu 
    +
    \obK
    \hbu = \b0 \, .
\end{equation}
\end{linenomath}
Unlike classical perturbation methods, which deploy the structural response as a function of the uncertain parameters by means of a Taylor expansion of the response about the null values of the uncertain parameters themselves, the improved perturbation method Taylor expands directly the linear momentum balance equation seeking for a better approximation of the response in terms of mean and variance \cite{impollonia:2002}. 

\subsection{Numerical solution strategy}
\label{sec:numerics}

For a large number of nodes (and trusses that carry the uncertain parameters), the direct solution of the two sets of equations \eqref{49}-\eqref{50} becomes prohibitive. Additionally, we note that the approach easily accounts for material nonlinearities, making the two sets of equations highly nonlinear. Therefore, we propose a solution strategy that can also be used for nonlinear and inelastic materials. We solve the two sets by adopting an iterative method based on alternate solutions. At the $r$ iteration, Eq.~\eqref{49} is solved to compute the average displacements $\bar \bu^{r+1}$, by assigning known values $\hat\bu^r_i$ to the variations $\hbu$ as
\begin{linenomath}
\begin{equation}\label{49bis}
    \left(   \obK
    +
    \frac{1}{2} 
    \bK_{ii} \langle \Theta_i^2 \rangle \right)  
    \bar \bu^{r+1}
    =
    \bF_{\rm ext} 
    - 
    \bK_i \hat\bu^r_i
    \langle \Theta_i^2 \rangle \, ,
\end{equation}
\end{linenomath}
while the $\hat\bu^{r+1}_i$ are obtained by solving the $N$ equations
\begin{linenomath}
\begin{equation}\label{50bis}
    \obK
    \hat\bu^{r+1}_i
    +
    \bK_i 
    \bar \bu^{r+1}
    = \b0 \, .
\end{equation}
\end{linenomath}

When convergence is achieved in the limit of an assigned tolerance on the displacement norm, from the resulting displacements we compute the mean value $\langle \bu \rangle$ and the covariance matrix $\Sigma_{\bu}$ 
\begin{linenomath}
\begin{equation}\label{51}
    \langle \bu \rangle = \obu \, \qquad \Sigma_{\bu}= \langle \bu \bu^T \rangle - \obu \obu^T= \langle \Theta_i^2\rangle \hbu \hbu^T.
\end{equation}
\end{linenomath}
The main diagonal terms of the covariance matrix $\Sigma_{\textbf{u}}$ provide important information on the stochastic model since each $[{\Sigma_\bu}]_{jj}$ represents the variance of the displacements associated with the j-th degree of freedom. The square root of these terms defines the standard deviation $\sigma_{jj}=\sqrt{[{\Sigma_\bu}]_{jj}}$, and denote the magnitude of the variability of the displacements with respect to the mean value $\bar u_j$. The Coefficient of Variation (CV) of the displacements field is defined as the ratio between the standard deviation and the corresponding average
\begin{linenomath}
$${\rm CV} = \sigma_{jj}/\bar u_j \, .$$
\end{linenomath} 

\begin{figure}[t]
	\centering
	\includegraphics[width=1\linewidth]{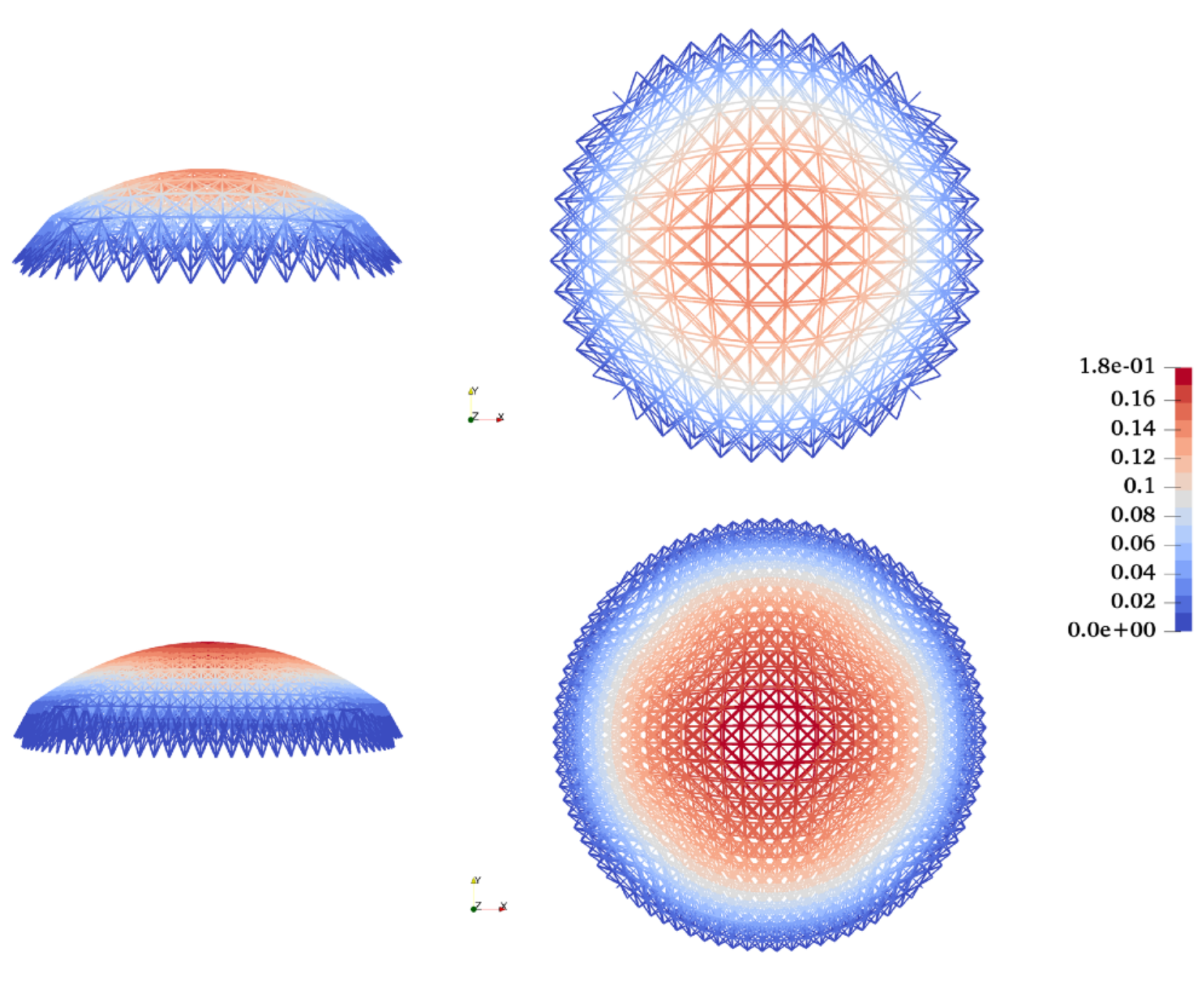} 
	\put(-500,200){\small{(a)}}
		\put(-300,200){\small{(b)}}
			\put(-500,0){\small{(c)}}
				\put(-300,0){\small{(d)}}
	\caption{Lateral and top views of mean values $\bar w$ of vertical displacements for two different meshes. Note the effect of more compliant radial trusses at the limbus.}
	\label{fig3:meanDispla}
\end{figure}
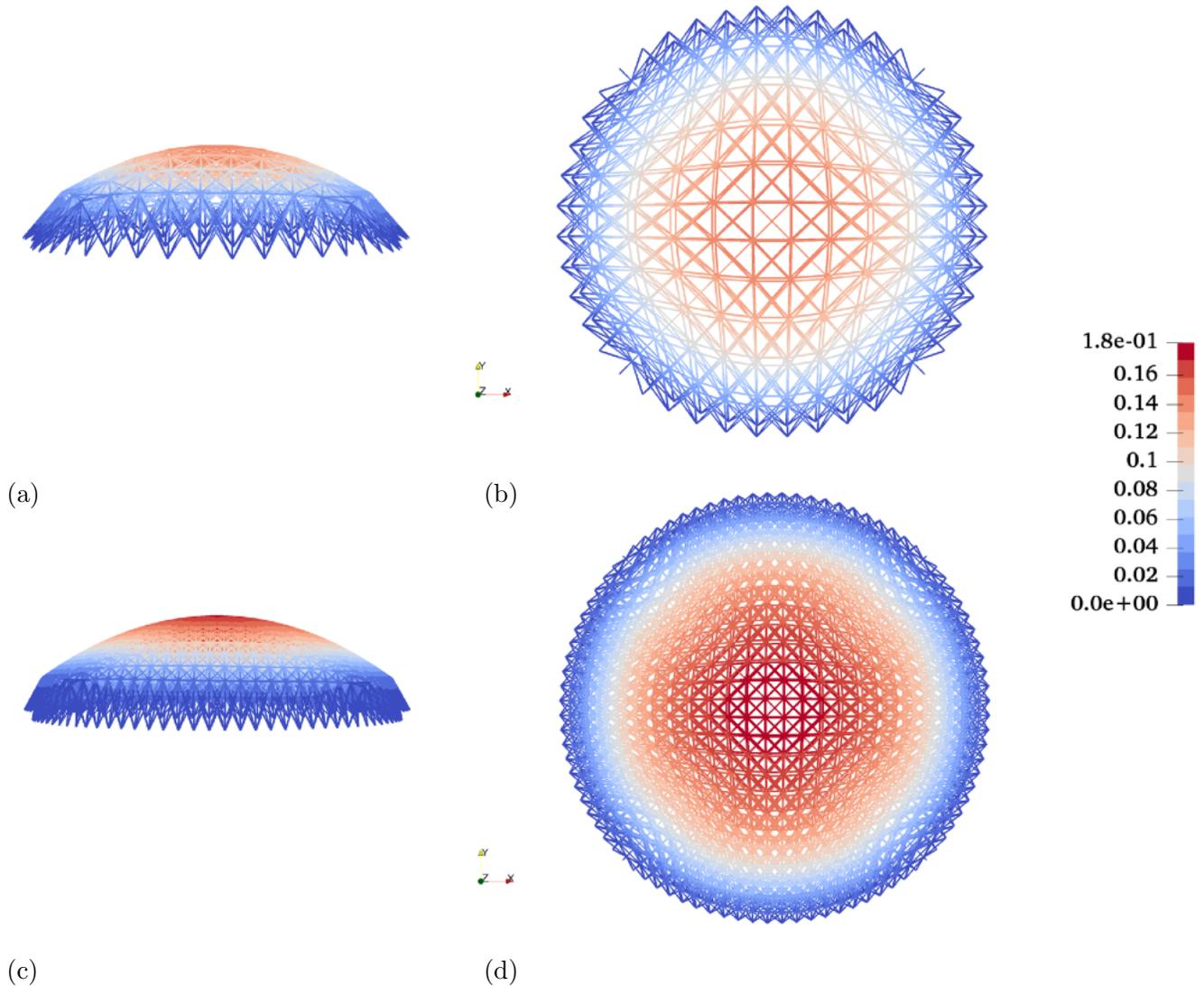

The assembled network that models the structure of the human cornea shown in Fig.~\ref{fig1:geom} includes standard trusses (crosslinks) and statistically enhanced trusses (collagen fibrils). A suitable model requires adopting a fine discretization that implies a large number of nodes and trusses, enlarging the size of the numerical problem ($N + 1$ systems of $3n$ equations) to very high levels. Implicit approaches that call for the inversion of the stiffness matrix become intractable from the computational point of view, making explicit methods more appealing. A robust solution strategy is given by the dynamic relaxation method,  proposed in \cite{OAKLEY199567} and optimized for treating truss structures in \cite{pandolfi2022modeling}. The idea behind dynamic relaxation is to achieve the solution of a static problem as the steady state solution of the equivalent pseudo-dynamic critically damped problem. The parameters of the pseudo-dynamics (fictitious mass and damping matrices) can be constructed to reach the fastest convergence toward the steady-state solution. 

\section{Results}
\label{sec4}

\begin{figure}[h]
	\centering	\includegraphics[width=1\linewidth]{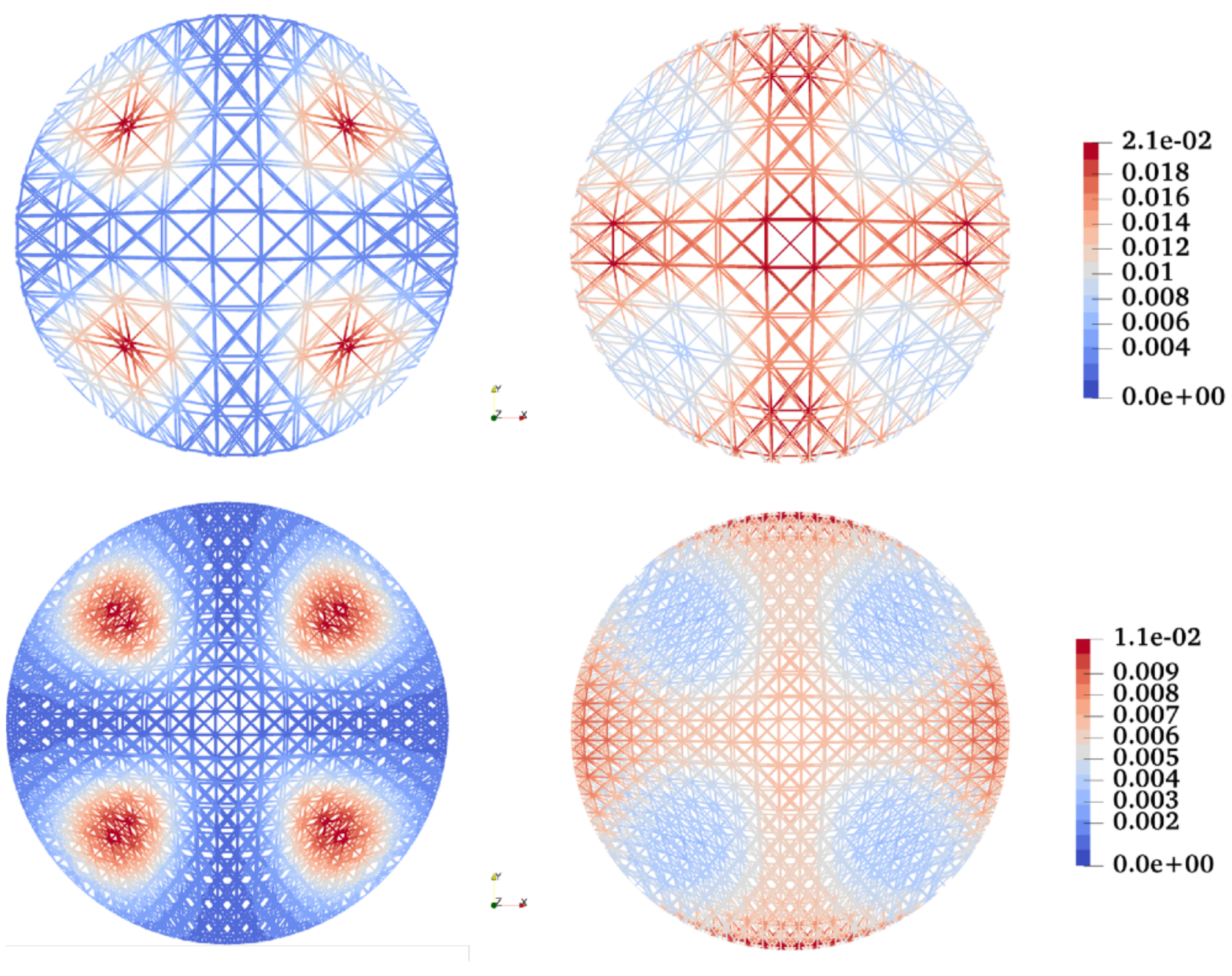} 
		\put(-500,200){\small{(a)}}
		\put(-300,200){\small{(b)}}
		\put(-500,0){\small{(c)}}
		\put(-300,0){\small{(d)}}
	\caption{(a,c) Distribution of the dispersion parameter $\kappa$ corresponding to the coarsest and finest meshes, respectively. (b,d) Standard deviation of the vertical displacement components $\sigma_{\hbu}$ corresponding to the coarsest and finest meshes, respectively.}
	\label{fig4:kappaSTD}
\end{figure}

We use the stochastic model to evaluate the mechanical response of the human cornea to the action of the IOP. We investigate how the randomness of the collagen fibril orientation affects displacement distribution and variability. We have considered different meshes obtained through uniform refinement so that any finer mesh is included in the coarser ones and verified that the mechanical response showed a convergent behavior. In the following, we show results for the mesh M1 (12 cells across the meridian section), consisting of 288 nodes and 3730 elements, and for the mesh M2 (24 cells across the meridian section), composed of 1152 nodes and 6724 elements. The most significant displacement component is the one in the direction of the optic axis (normal to the corneal surfaces at the apex). This displacement component is denoted with $w$.  

We begin with the distribution of the concentration parameter $b$, see Fig.~\ref{fig2:vonMises}(b). Accordingly to experimental observations, the model shows the maximum dispersion localized in four lobes separated by the strong fibril alignments in the NT and SI directions. Strong alignment is also observed at the limbus.

Figure~\ref{fig3:meanDispla} compares, in terms of color maps, the mean value $\bar w$ for the coarser and the finer meshes at the physiological IOP (assumed to be 14 mmHg). The images show the side and top views. The displacements at the corneal apex are in the range of the values obtained with continuum models \cite{pandolfi:2006}; as expected, the coarse mesh provides a stiffer response. Results show that the model does not perform well in the proximity of the limbus because the truss model cannot reproduce the continuity of the constraint offered by the limbus. For the particular architecture of the model, the radial trusses at the limbus deform more because their axial forces are overestimated. Keeping in mind this drawback, we restrict our observation to the central part of the cornea, that we can identify with the optical zone (where the light rays deviate to the retina and the area of primary interest in refractive surgery). We estimate the optical zone as the central spherical cap with 4 mm in-plane radius.

Figure~\ref{fig4:kappaSTD}(a,c) shows the top view of the contour maps of the dispersion parameter $\kappa$ for the coarse and the fine meshes. The dispersion parameter ranges in the interval $[0,1/3]$. The $\kappa$ values in Fig.~\ref{fig4:kappaSTD} range in a more restricted interval because the visualization refers to the averaged values at the nodes, which also accounts for the deterministic trusses for which $\kappa=0$. The standard deviation plots in Fig.~\ref{fig4:kappaSTD}(b,d) refer to the sole component $w$ of the displacements. The images reveal that, along NT and SI directions, where the collagen fibers are more aligned, and the overall response is expected to be more anisotropic, the standard deviation is characterized by higher values. Contrariwise, in the four zones of maximum fibril dispersion where a pseudo-isotropic response is expected, the standard deviation is characterized by low values, close to zero, confirming the mean value of the displacements $\bar w$ is perfectly representative of the actual stochastic displacements in those areas.
\begin{figure}[h]	
\centering
\includegraphics[width=\textwidth]{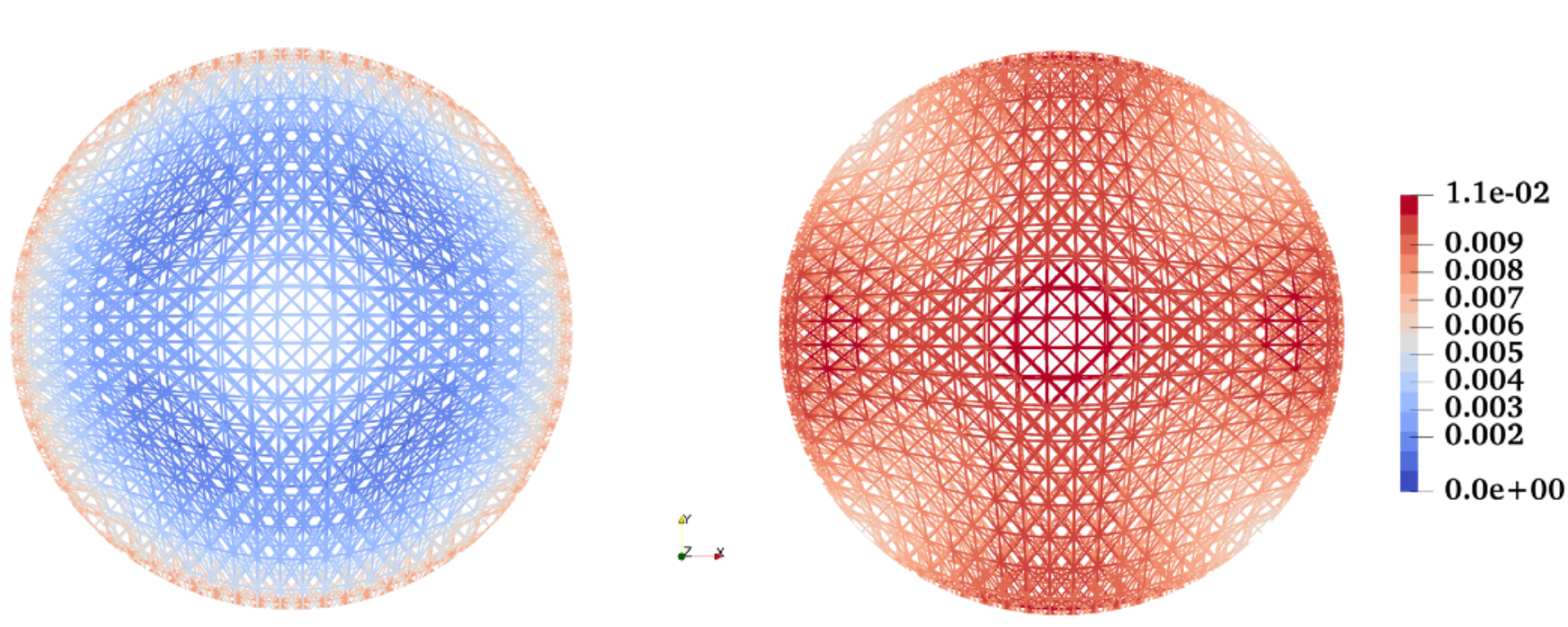}
	\put(-500,0){\small{(a)}}
		\put(-300,0){\small{(b)}}
\caption{Standard deviation of the vertical displacement components $\sigma_{\hbu}$ for the finest mesh corresponding to a uniform distribution of the concentration parameter. (a) $b=0$, (b) $b=8$.}
	\label{fig5:StandardDev}
\end{figure}
\begin{figure}[h]	
\centering
\includegraphics[width=0.78\textwidth]{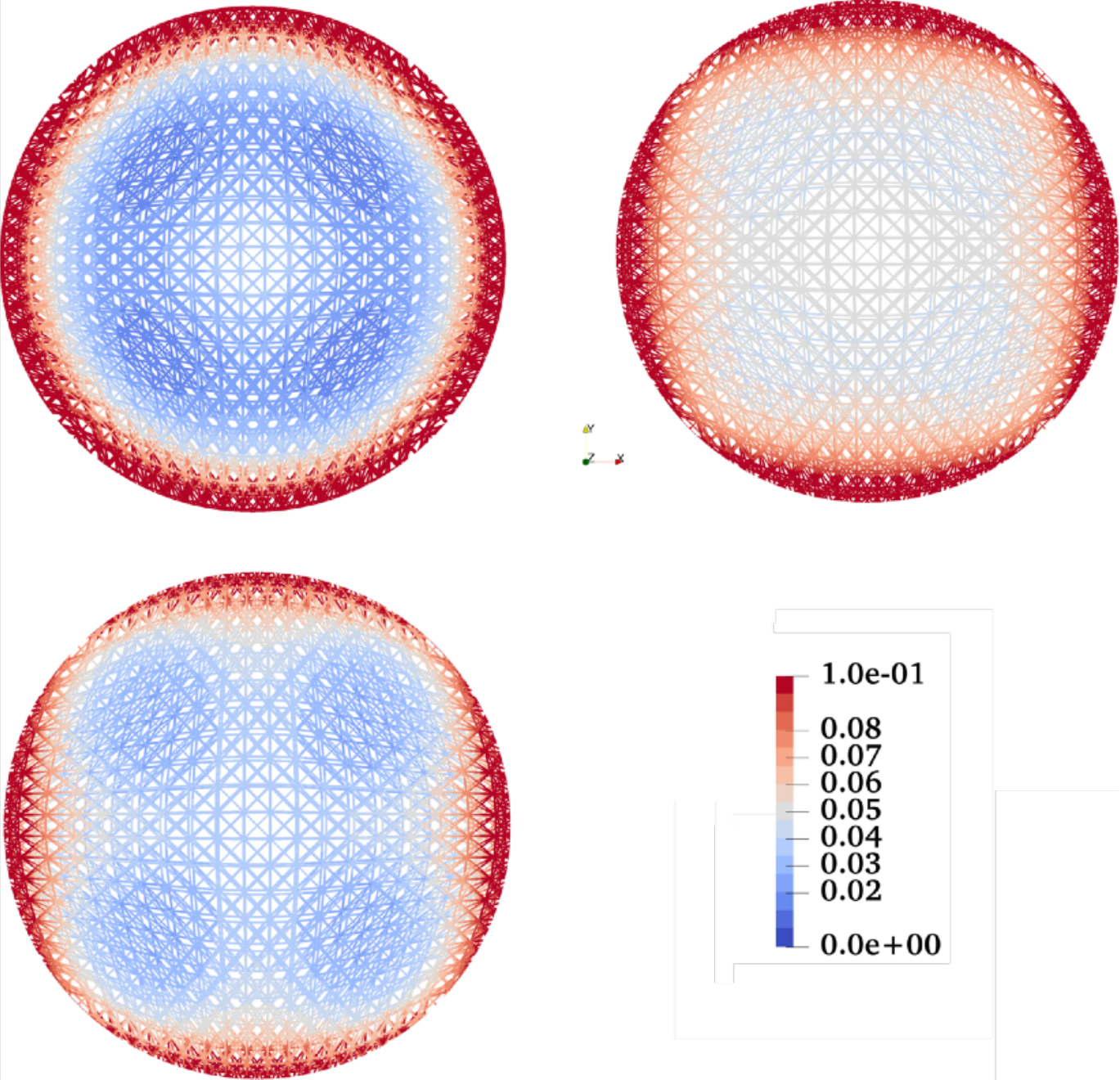}
	\put(-400,200){\small{(a)}}
		\put(-160,200){\small{(b)}}
			\put(-400,0){\small{(c)}}
\caption{Coefficient of variation CV for the coarsest mesh related to the vertical displacement component. (a) $b=0$ uniform, (b) $b=8$ uniform, (c) $b$ variable as in Figure \ref{fig2:vonMises}.}
	\label{fig6:VariationCoeff}
\end{figure}

For the sake of comparison and to better understand how the concentration parameter $b$ affects the randomness of the mechanical response of the model, Fig.~\ref{fig5:StandardDev} shows the contour levels of the standard deviation of the $w$ displacement for two idealized extreme cases where the concentration parameter is assumed to be uniform. In the first case, we assume $b=0$ ($\kappa=1/3$), which corresponds to the case of fully dispersed fibrils and isotropic behavior for the cornea shell, Fig.~\ref{fig5:StandardDev}(a). In the second case, we assume  $b=8$ ($\kappa \simeq 0$), which corresponds to the case of a bunch of fibrils strongly aligned in the mean direction ($\ba_0$) of the distribution, defining an anisotropic (orthotropic) behavior for the cornea shell, Fig.~\ref{fig5:StandardDev}(b). Both cases indicate an almost uniform distribution of the standard deviation, the isotropic case being characterized by lower standard deviation values. The isotropic case results are more sensitive to the presence of the boundary, where the model, as already explained, does not reflect the behavior of the limbus. 

These results confirmed that the standard deviation keeps low values for a very dispersed collagen fibril distribution. At the same time, it reaches higher uniform values when the alignment of the collagen fibrils induces an anisotropic behavior. 
 
Fig.~\ref{fig6:VariationCoeff} visualizes the contour levels of the CV relative to the sole displacement $w$ for the three cases considered, i.~e., $b=0$, $b=8$, and $b$ according to the distribution described in Fig.~\ref{fig2:vonMises}(c). The maximum values of the standard deviation observed in the optic zone reach about 10$\%$ of the corresponding mean values. In all cases, $\bar w$ remains relatively uniform in the central zone, decreasing to zero in the radial directions toward the limbus. The trend suggests that, in the central zone, the CV follows the distribution of the standard deviation $\sigma_{ii}$ closely, while it rapidly increases in the annulus approaching the limbus.

\section{Discussion and conclusion}
\label{sec:fR}

By using a discrete model of the cornea that accounts for the microstructural components of the stroma according to their physical function, we wanted to characterize the mechanical properties of the collagen fibrils, which play the role of structural elements, and to evaluate the effect of uncertainties on the displacement fields. We considered only the cornea response to the physiological IOP, to validate a model that in the future can be extended and used for more complex simulations, connected to refractive surgery or cornea degeneration. 

The results presented here confirm that the statistical nature of the fibril distribution has a remarkable impact on the displacements of the cornea under the action of the (deterministic) IOP and over an assigned geometry of the model. More specifically, results show that the maximum variability with respect to the mean displacements is found in the optical zone, in both NT and SI directions. This observation may be of interest to refractive surgeons, since refractive surgery procedures are actually performed over the zones where, in the mechanical characterization, the maximum values of uncertainty are observed.

Considering the biological variability of the cornea among individuals, a more predictive model must include not only the patient-specific geometry, with all the relevant uncertainties, but also the IOP, which is not a constant load and is generally unknown. IOP varies among patients and fluctuates in the hour timescale; therefore, it should also be characterized stochastically.

Based on improved perturbation analysis, the adopted method has been validated by comparing the stochastic results with the results of the corresponding deterministic model \cite{pandolfi:2019}. The reliability of the model has been assessed by considering the stochastic response in the limit cases of isotropic and strongly anisotropic (orthotropic) materials. We remark that the solution of complex stochastic problems has been possible by adopting fully explicit solution approaches (dynamic relaxation), particularly suitable for concurrent computing. 

The encouraging results suggest the pathway for future developments, which include the extension of the model to the cases of geometric or mechanical nonlinearities, which have been already considered by a few authors in deterministic numerical models. As matter of facts, experimental evidence suggests that large deformation phenomena occur in the cornea. Furthermore, corneal diseases related to a loss of the mechanical performance, such as keratoconus, require the introduction of degenerative effects in the numerical model. In both cases, a sound treatment of all sort of uncertainties potentially affecting the mechanical parameters could improve the characterization of the structural response.

\section*{Acknowledgements}
The research has been developed under the auspices of the Italian National Group of Physics-Mathematics (GNFM) of the Italian National Institution of High Mathematics ‘‘Francesco Severi’’ (INDAM).

\section*{Conflict of interest}
The authors declare that they have no conflict of interest.

%\bibliography{statisticalTruss}

\newpage
\bibliographystyle{elsarticle-num}
\bibliography{statisticalTruss}

\begin{thebibliography}{10}
\expandafter\ifx\csname url\endcsname\relax
  \def\url#1{\texttt{#1}}\fi
\expandafter\ifx\csname urlprefix\endcsname\relax\def\urlprefix{URL }\fi
\expandafter\ifx\csname href\endcsname\relax
  \def\href#1#2{#2} \def\path#1{#1}\fi

\bibitem{meek:2009}
K.~M. Meek, C.~Boote, The use of {X}-ray scattering techniques to quantify the
  orientation and distribution of collagen in the corneal stroma, {Progress in
  retinal and eye research} 28~(5) (2009) 369--392.

\bibitem{kokott1938mechanisch}
W.~Kokott, {\"U}ber mechanisch-funktionelle strukturen des auges, Albrecht von
  Graefes Archiv f{\"u}r Ophthalmologie 138~(4) (1938) 424--485.

\bibitem{pandolfi:2019}
A.~Pandolfi, A.~Gizzi, M.~Vasta, A microstructural model of cross-link
  interaction between collagen fibrils in the human cornea, Philosophical
  Transactions of the Royal Society A 377 (2019) 20180079.

\bibitem{Gizzi:2018aa}
A.~Gizzi, A.~Pandolfi, M.~Vasta,
  \href{https://doi.org/10.1007/s10665-017-9943-5}{A generalized statistical
  approach for modeling fiber-reinforced materials}, Journal of Engineering
  Mathematics 109~(1) (2018) 211--226.
\newblock \href {https://doi.org/10.1007/s10665-017-9943-5}
  {\path{doi:10.1007/s10665-017-9943-5}}.
\newline\urlprefix\url{https://doi.org/10.1007/s10665-017-9943-5}

\bibitem{pandolfi2022modeling}
A.~Pandolfi, M.~L. De~Bellis, A.~Gizzi, M.~Vasta, Modeling the degeneration of
  the collagen architecture in a microstructural model of the human cornea,
  Mathematics and Mechanics of Solids (2022) 10812865221092690.

\bibitem{stefanou2009stochastic}
G.~Stefanou, The stochastic finite element method: past, present and future,
  Computer methods in applied mechanics and engineering 198~(9-12) (2009)
  1031--1051.

\bibitem{pryse2018projection}
S.~E. Pryse, A.~Kundu, S.~Adhikari, Projection methods for stochastic dynamic
  systems: A frequency domain approach, Computer Methods in Applied Mechanics
  and Engineering 338 (2018) 412--439.

\bibitem{falsone2013explicit}
G.~Falsone, D.~Settineri, Explicit solutions for the response probability
  density function of linear systems subjected to random static loads,
  Probabilistic Engineering Mechanics 33 (2013) 86--94.

\bibitem{navarro2022new}
A.~Navarro-Quiles, R.~Laudani, G.~Falsone, A new stochastic method based on the
  taylor expansion to compute response probability densities of uncertain
  systems, International Journal for Numerical Methods in Engineering (2022).

\bibitem{ghanem2003stochastic}
R.~G. Ghanem, P.~D. Spanos, Stochastic finite elements: a spectral approach,
  Courier Corporation, 2003.

\bibitem{liu1986probabilistic}
W.~K. Liu, T.~Belytschko, A.~Mani, Probabilistic finite elements for nonlinear
  structural dynamics, Computer Methods in Applied Mechanics and Engineering
  56~(1) (1986) 61--81.

\bibitem{kleiber1992stochastic}
M.~Kleiber, T.~D. Hien, The stochastic finite element method: basic
  perturbation technique and computer implementation, Wiley, 1992.

\bibitem{elishakoff1995improved}
I.~Elishakoff, Y.~Ren, M.~Shinozuka, Improved finite element method for
  stochastic problems, Chaos, Solitons \& Fractals 5~(5) (1995) 833--846.

\bibitem{muscolino2000improved}
G.~Muscolino, G.~Ricciardi, N.~Impollonia, Improved dynamic analysis of
  structures with mechanical uncertainties under deterministic input,
  Probabilistic Engineering Mechanics 15~(2) (2000) 199--212.

\bibitem{impollonia2002static}
N.~Impollonia, G.~Muscolino, Static and dynamic analysis of non-linear
  uncertain structures, Meccanica 37~(1) (2002) 179--192.

\bibitem{van2003modal}
B.~Van~den Nieuwenhof, J.-P. Coyette, Modal approaches for the stochastic
  finite element analysis of structures with material and geometric
  uncertainties, Computer Methods in Applied Mechanics and Engineering
  192~(33-34) (2003) 3705--3729.

\bibitem{kaminski2013stochastic}
M.~Kaminski, The stochastic perturbation method for computational mechanics,
  John Wiley \& Sons, 2013.

\bibitem{kasinos2021reduced}
S.~Kasinos, A.~Palmeri, M.~T. Lombardo, S.~Adhikari, A reduced modal subspace
  approach for damped stochastic dynamic systems, Computers \& Structures 257
  (2021) 106651.

\bibitem{koery:2023}
J.~Koery, N.~Hill, Z.~Y. Luo, P.~S. Stewart, A.~Pandolfi, A two-dimensional
  model of the degeneration of the carrying structure of the cornea: upscaling
  from discrete to continuum, (in preparation) (2023).

\bibitem{marino:2013}
M.~Marino, G.~Vairo, Multiscale elastic models of collagen bio-structures: From
  cross-linked molecules to soft tissues, Springer Berlin Heidelberg, Berlin,
  Heidelberg, 2013, pp. 73--102.

\bibitem{gyi:1988}
T.~Gyi, K.~M. Meek, G.~F. Elliott, Collagen interfibrillar distances in corneal
  stroma using synchrotron x-ray diffraction: a species study, International
  Journal of Biological Macromolecules 10 (1988) 265--269.

\bibitem{scott:2003}
J.~E. Scott, Elasticity in extracellular matrix 'shape modules' of tendos,
  cartilage, etc. a sliding proteoglycan-filament model, Journal of Physics
  553~(2) (2003) 335--343.

\bibitem{pandolfi2012fiber}
A.~Pandolfi, M.~Vasta, Fiber distributed hyperelastic modeling of biological
  tissues, Mechanics of Materials 44 (2012) 151--162.

\bibitem{vasta2014three}
M.~Vasta, A.~Gizzi, A.~Pandolfi, On three-and two-dimensional fiber distributed
  models of biological tissues, Probabilistic Engineering Mechanics 37 (2014)
  170--179.

\bibitem{impollonia:2002}
N.~Impollonia, G.~Muscolino, Static and dynamic analysis of non-linear
  uncertain structures, Meccanica 37~(1) (2002) 179--192.

\bibitem{OAKLEY199567}
D.~R. Oakley, N.~F. Knight,
  \href{https://www.sciencedirect.com/science/article/pii/004578259500805B}{Adaptive
  dynamic relaxation algorithm for non-linear hyperelastic structures part i.
  formulation}, Computer Methods in Applied Mechanics and Engineering 126~(1)
  (1995) 67--89.
\newblock \href {https://doi.org/https://doi.org/10.1016/0045-7825(95)00805-B}
  {\path{doi:https://doi.org/10.1016/0045-7825(95)00805-B}}.
\newline\urlprefix\url{https://www.sciencedirect.com/science/article/pii/004578259500805B}

\bibitem{pandolfi:2006}
A.~Pandolfi, F.~Manganiello, A material model for the human cornea,
  {Biomechanics and Modelling in Mechanobiology} 5 (2006) 237--246.

\end{thebibliography}

\end{document}